\documentclass[aps,prl,amsmath,amssymb,reprint,superscriptaddress,showpacs]{revtex4-1}
\usepackage{graphicx}
\usepackage{ulem}

\usepackage{color}

\begin{document}

\title{Neutron-skin thickness of  $^{208}$Pb, and symmetry-energy constraints
 from the study  of the anti-analog giant  dipole resonance}

\author{A. Krasznahorkay},
\email[]{kraszna@atomki.hu}
\affiliation{Inst. for Nucl. Res. (MTA-Atomki), H-4001 Debrecen, P.O. Box 51, Hungary}
\author{M. Csatl\'os}
\affiliation{Inst. for Nucl. Res. (MTA-Atomki), H-4001 Debrecen, P.O. Box 51, Hungary}
\author{L. Csige}
\affiliation{Inst. for Nucl. Res. (MTA-Atomki), H-4001 Debrecen, P.O. Box 51, Hungary}
\author{T.K. Eriksen}
\address{Department of Physics, University of Oslo, N-0316 Oslo, Norway}
\author{F. Giacoppo}
\address{Department of Physics, University of Oslo, N-0316 Oslo, Norway}
\author{A. G\"orgen}
\address{Department of Physics, University of Oslo, N-0316 Oslo, Norway}
\author{T.W. Hagen}
\address{Department of Physics, University of Oslo, N-0316 Oslo, Norway}
\author{M.N.~Harakeh}
\address{KVI, University of Groningen, Groningen, The Netherlands}
\address{GANIL, CEA/DSM-CNRS/IN2P3, 14076 Caen, France}
\author{R. Julin}
\address{Department of Physics, University of Jyv\"skyl\"a, Jyv\"skyl\"a, FIN-40014, Finland}
\author{P. Koehler}
\address{Department of Physics, University of Oslo, N-0316 Oslo, Norway}
\author{N. Paar}
\address{Physics Department, Faculty of Science, University of Zagreb, Croatia}
\author{S. Siem}
\address{Department of Physics, University of Oslo, N-0316 Oslo, Norway}
\author{L. Stuhl}
\affiliation{Inst. for Nucl. Res. (MTA-Atomki), H-4001 Debrecen, P.O. Box 51, Hungary}
\author{T. Tornyi}
\affiliation{Inst. for Nucl. Res. (MTA-Atomki), H-4001 Debrecen, P.O. Box 51, Hungary}
\author{D. Vretenar}
\address{Physics Department, Faculty of Science, University of Zagreb, Croatia}

\begin{abstract}
The $^{208}$Pb($p$,$n\gamma\bar p$) $^{207}$Pb reaction at a beam energy of 30
MeV has been used to excite the anti-analog of the giant dipole resonance (AGDR) 
and to measure its $\gamma$-decay to the isobaric analog state in coincidence with proton decay of IAS. 
The energy of the transition has also been calculated with the 
self-consistent relativistic random-phase approximation (RRPA), and found to be linearly 
correlated to the predicted value of the neutron-skin thickness ($\Delta R_{pn}$). By comparing 
the theoretical results with the measured transition energy, the value 
of 0.190 $\pm$ 0.028 fm has been determined for $\Delta R_{pn}$ of
$^{208}$Pb, in agreement with previous experimental results. The AGDR 
excitation energy has also been used to calculate the symmetry energy
at saturation ($J=32.7 \pm 0.6$ MeV)  and the slope of the symmetry energy ($L=49.7 \pm 4.4$ MeV), resulting in more stringent constraints than most of the previous studies. 
\end{abstract}


\maketitle

\section{Introduction}
There is a renewed interest in measuring  the
thickness of the neutron skin \cite{ab12,Pie12,ro11,ta11}, because it
constrains the symmetry-energy term of the nuclear equation of
state. The precise knowledge of the symmetry energy is essential not
only for describing the structure of neutron-rich nuclei, but also for
describing the properties of the neutron-rich matter in nuclear
astrophysics.

The symmetry energy determines to a large extent, through the Equation
of State (EoS), the proton fraction of neutron stars \cite{la01}, the
neutron skin in heavy nuclei \cite{fu02} and enters as input in the
analysis of heavy-ion reactions \cite{li98, ba02}. Furnstahl
\cite{fu02} demonstrated that in heavy nuclei an almost linear
empirical correlation exists between the neutron-skin thickness and
theoretical predictions for the symmetry energy of the EoS in terms of
various mean-field approaches. This observation has contributed to a
revival of an accurate determination of the neutron-skin thickness in
neutron-rich nuclei \cite{te08,ta11,ro11,ab12}.  In this work, we
suggest a new method for measuring the neutron-skin thickness with
unprecedented accuracy.

Recently, we have shown that the energy difference between the
anti-analog giant dipole resonance (AGDR) and the isobaric analog
state (IAS) is very sensitively related to the corresponding
neutron-skin thickness \cite{kr13}. We have also calculated the energy
of the AGDR for the $^{208}$Pb isotope using the state-of-the-art
fully self-consistent relativistic proton-neutron quasi-particle
random-phase approximation and compared to the available experimental
data after correcting them for the admixture of the isovector spin giant dipole
resonance (IVSGDR) \cite{kr13a}.

Yasuda \textit{et al.} \cite{ya13} separated the AGDR from other
excitations, such as the IVSGDR, by multipole decomposition analysis of
the $^{208}$Pb($\vec p, \vec n$) reaction at a bombarding energy of
E$_p=$296 MeV. The polarization transfer observables were found to be
useful for carrying out this separation. The energy difference between
the AGDR and the IAS was determined to be
$\Delta$E$_{AGDR-IAS}$ = 8.69 $\pm$ 0.36 MeV, where the uncertainty
includes both statistical and systematic contributions.  Using our
theoretical results \cite{kr13a} a neutron-skin thickness of $\Delta
R_{pn}$ = 0.216 $\pm$ 0.046 $\pm$ 0.015 fm could be obtained, where
the first and second uncertainties are the experimental and
theoretical one, respectively.

The aim of the present work is to determine $\Delta$E$_{AGDR-IAS}$
with high precision by measuring the energy of the corresponding
$\gamma$-transition. The direct $\gamma$-branching ratio of the AGDR
to the IAS is expected to be similar to that of the isovector giant
dipole resonance (IVGDR) to the ground-state (g.s.) in the parent
nucleus, which can be calculated from the parameters of the IVGDR
\cite{kr94}.

\section{The anti-analog giant dipole resonance and its $\gamma$ decay}

Due to the isovector nature of the ($p$,$n$) reaction, the strength of
the E1 excitation is distributed into T$_0$-1, T$_0$ and T$_0$+1
components, where T$_0$ is the g.s. isospin of the
initial nucleus.  The relevant Clebsch-Gordan coefficients
\cite{os92} show, that the T$_0$-1 component (AGDR) is favored compared to the
T$_0$ and T$_0$+1 ones by factors of about T$_0$, and 2T$_0^2$,
respectively. According to the work of Osterfeld \cite{os92}, the
non-spin-flip transition is preferred at low bombarding energies below 50 MeV.

Dipole resonances were excited earlier at such low energies in the
$^{208}$Pb($p$,$n$) reaction by Sterrenburg  \textit{et al.} \cite{st80}, and
Nishihara \textit{et al.} \cite{ni85} at E$_p$= 45 MeV and 41 MeV,
respectively.  However, it was shown experimentally \cite{os81,au01}
that the observed $\Delta L$= 1 resonance was a superposition of all
possible IVSGDR modes and the non-spin-flip dipole AGDR even at these low
bombarding energies.

\begin{figure}[bht]
\centering
\includegraphics[width=80mm]{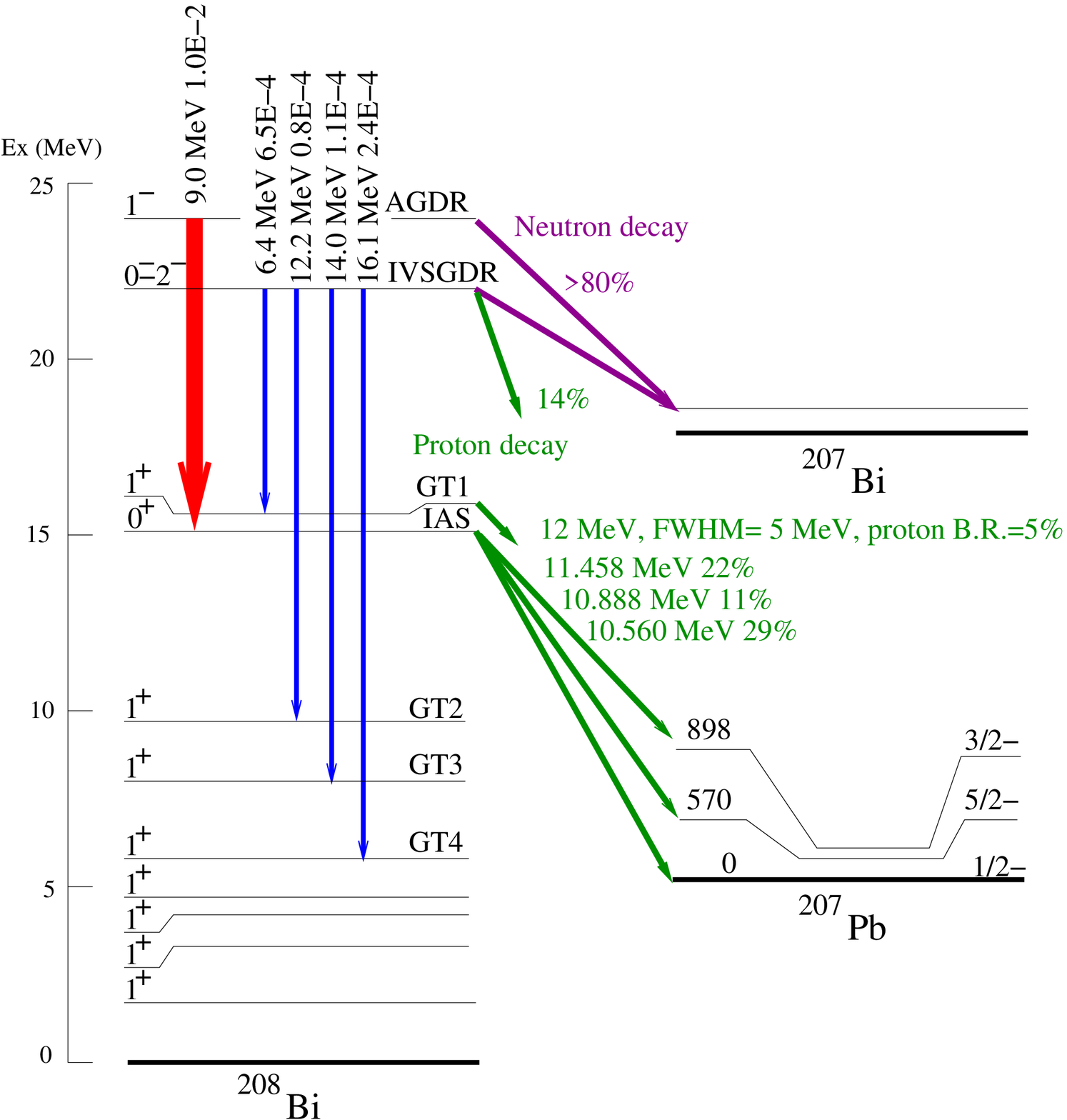}
\caption{\label{fig:en_lev}Energy levels excited in the $^{208}$Pb($p$,$n$)$^{208}$Bi
  reaction and their expected $\gamma$-decay branching ratios (red
  and blue  arrows). The energies and branching ratios of the
  proton decay of the IAS and GT resonance to the low-lying states in $^{207}$Pb are also
  shown (green arrows).
\label{AGDR}}
\end{figure}

The expected $\gamma$-decay properties of the states excited in
$^{208}$Bi are shown in Fig.~\ref{fig:en_lev} together with the proton-decay branching
ratios of the IAS \cite{igo69,cr72,bh77}.

The observed $\gamma$-ray branching ratio of the IVGDR to the g.s. of
$^{208}$Pb is about 1\% \cite{kr94}. In contrast, in the investigation
of the electromagnetic decay properties of the IVSGDR to the low-lying
Gamow-Teller (GT) states by Rodin and Dieperink \cite{ro02} the
$\gamma$-decay branching ratio was found in the range of 10$^{-4}$.

\section{Experimental methods and results}

The experiment, aiming at studying the neutron-skin thickness
of $^{208}$Pb, was performed at the Oslo Cyclotron Laboratory (OCL)
with 30 MeV proton beam bombarding a 5.5-mg/cm$^2$ thick, self-supporting metallic $^{208}$Pb
target and a 1 mg/cm$^2$ thick $^{12}$C
target for energy calibration.

In the experiment, the proton-decay of the IAS was used as a signature
of the de-excitation of the IAS. The $\gamma$-transition from the
decay of the AGDR was measured in coincidence with such proton lines.
These particle-$\gamma$ coincidences were measured with the SiRi
particle telescope and CACTUS $\gamma$-detector systems
\cite{gu11,gu90}. The SiRi detectors were placed at backward
angles, covering an angular range of $\Theta$=126$^\circ$-140$^\circ$
relative to the beam axis. The $\Delta$E and E detectors had
thicknesses of 130 $\mu$m and 1550 $\mu$m, respectively. The CACTUS
array consists of 28 collimated 5''$\times$ 5'' NaI(Tl) detectors with
a total efficiency of 15.2\% for $E_\gamma$= 1.33 MeV.

\begin{figure}[bht]
\centering
\includegraphics[width=80mm]{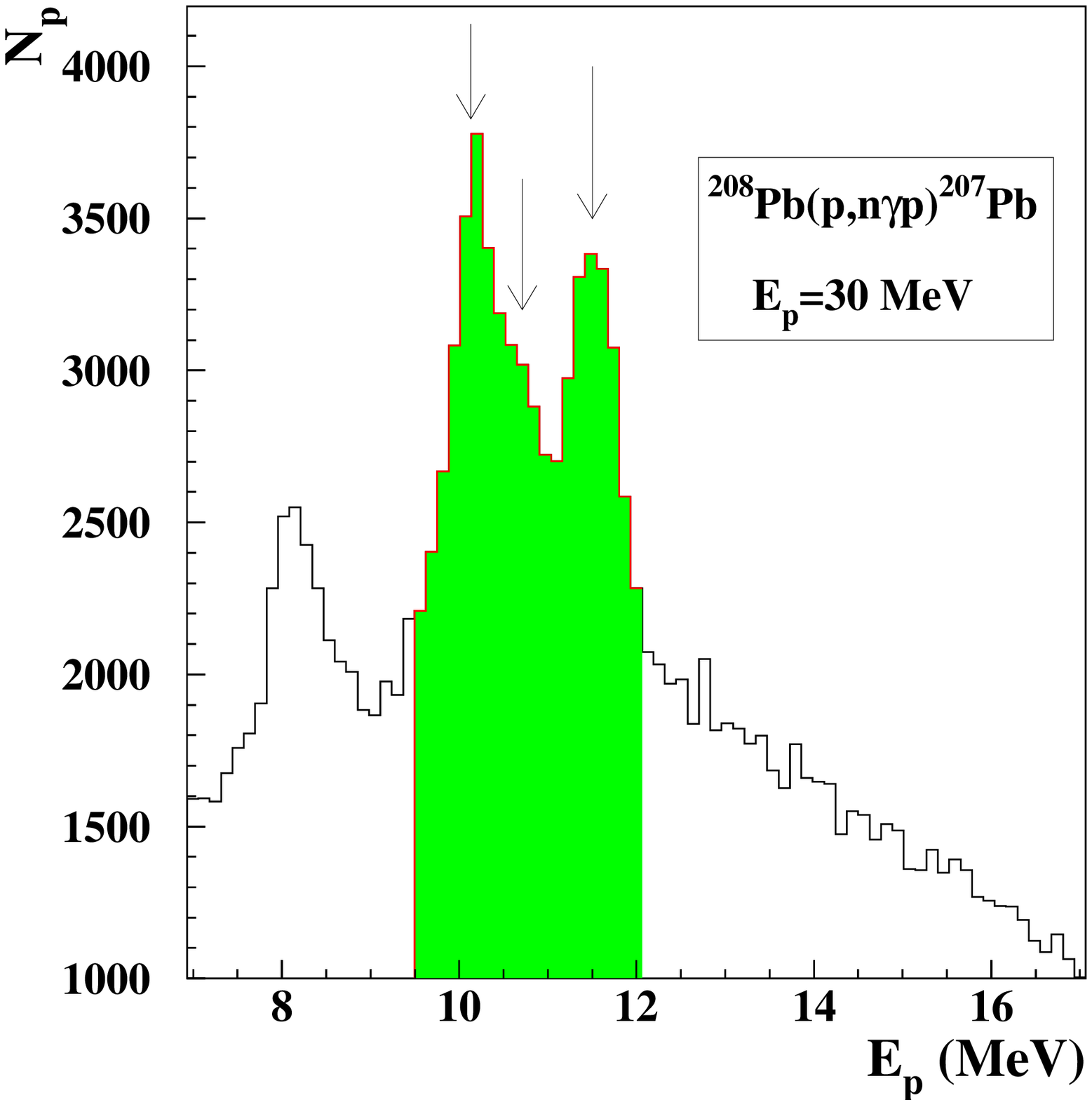}
\caption{\label{fig:proton}Proton energy spectrum measured in coincidence with the
$\gamma$ rays.}
\end{figure}

A typical proton spectrum is shown in Fig.~\ref{fig:proton}. The
proton transitions populating the low-lying states in $^{207}$Pb are
marked by arrows and used for gating the $\gamma$ rays.

\begin{figure}[bht]
\centering
\includegraphics[width=80mm]{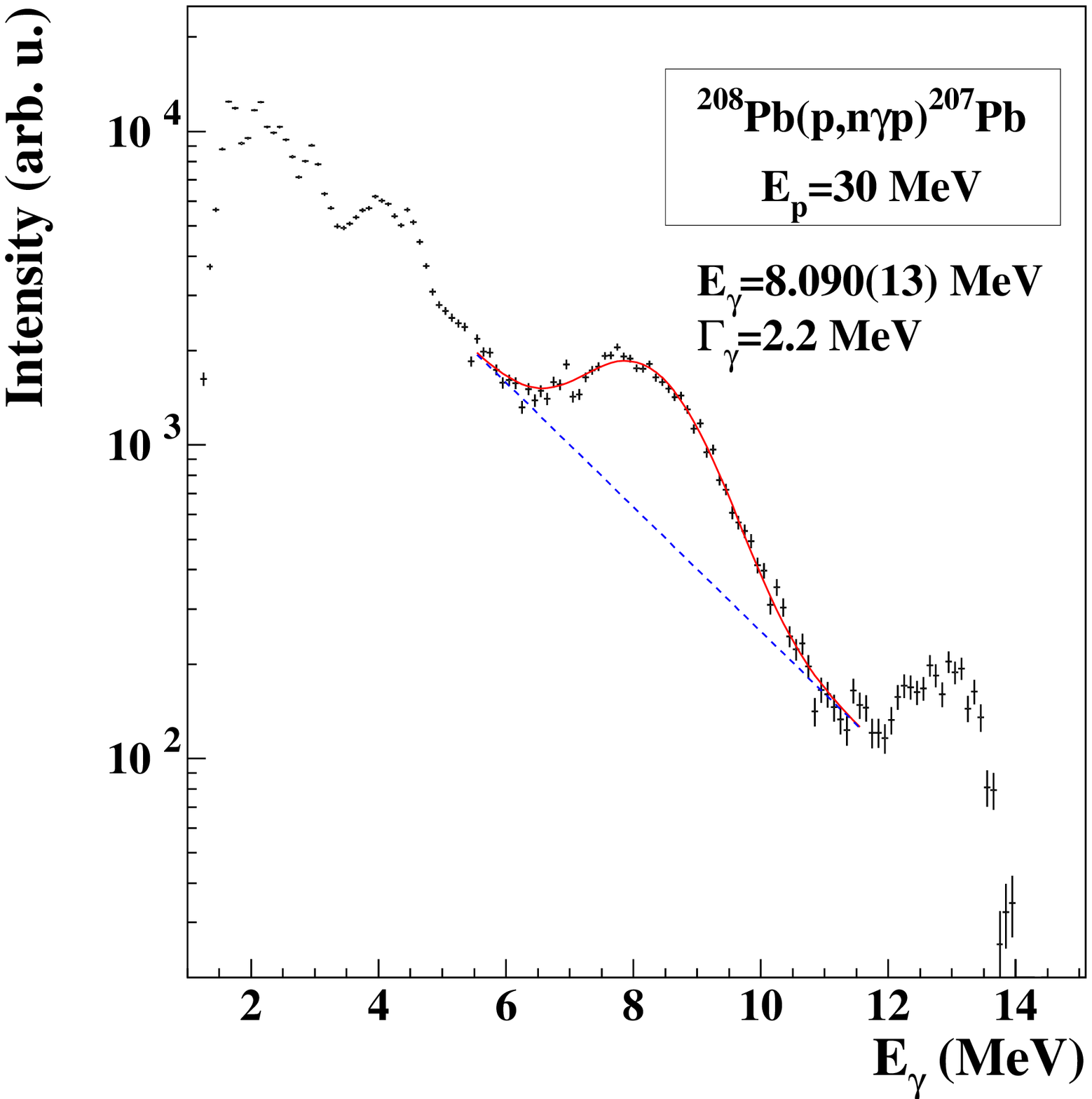}
\caption{\label{fig:gamma}The $\gamma$-ray energy spectrum measured in coincidence with
  protons of energy $9.5 \leq E_p\leq 12$ MeV. The random coincidences
  were subtracted and the spectrum was corrected for the efficiency of
  the NaI detectors. The solid line shows the result of the fit
  described in the text.}
\end{figure}

The energy of the $\gamma$ rays was measured in coincidence with the
protons stemming from the decay of the IAS in $^{208}$Bi. The random
coincidence contribution was subtracted as well as the contribution of
the proton decay of the GTR, which represents a broad
($\Gamma\approx$2.9 MeV) background in the proton spectrum. 

The
centroid of the $\gamma$ transition was shifted towards lower energies as a
result of the decreasing efficiency of the NaI detectors. In order to
correct this effect, the spectrum was normalized with the detector
response function that was extracted experimentally in
Refs.~\cite{gu11,gu90}. The $\gamma$-ray energy spectrum, as a result
of these corrections, is presented in Fig.~\ref{fig:gamma} together
with the statistical error bars.

The double line at 4.44 MeV comes from carbon contamination of the
target excited in the ($p$,$p\prime$) reaction, while the broad
transition around 13.3 MeV may come from the decay of the IVGDR
excited in $^{208}$Pb by the same reaction. As the IVGDR is broad
($\Gamma$=3.6 MeV) the inelastically scattered protons should have a
broad distribution. Unfortunately, the $\gamma$-ray spectrum does not
cover the full energy region of the IVGDR.

Additionally, NaI detectors are sensitive to low-energy neutrons
\cite{ha83}. These are captured mostly by iodine and the
$^{127}$I($n$,$\gamma$) reaction produces $\gamma$-rays with an energy
of E$_{\gamma}$=6.826 MeV, which interfere with the low-energy
side of the AGDR $\rightarrow$ IAS transition. At higher neutron
energies the neutron-capture cross section decreases drastically,
and the response of the NaI detectors for MeV neutrons is constant
as a function of energy.

The NaI detectors of the CACTUS setup were placed relatively close
(d=22 cm) to the target. Therefore, the time-of-flight method could not be used
to discriminate safely against neutrons produced in the
$^{208}$Pb($p$,$n$) reaction and also in the decay of the giant
resonances. The effect of these neutrons had to be carefully
treated. On the other hand, according to previous experimental studies
\cite{st80, ni85}, neutrons from the $^{208}$Pb($p$,$n$) reaction are
ejected predominantly to forward directions, and the cross section of
this reaction drops by one order of magnitude beyond 30 degrees. Since
the smallest angle of the NaI detectors of the CACTUS setup was
39$^\circ$ with respect to the beam direction, the ejected neutrons
did not disturb the $\gamma$-spectrum considerably.

Giant resonances (including the AGDR) decay also by neutrons, which
are detected by CACTUS with high efficiency. However, such neutron
emission goes to the low-lying states of $^{207}$Bi, and therefore
such neutrons are not in coincidence with the proton-decay of the IAS
in $^{208}$Bi. These neutrons contributed to the random coincidences
only, which were subtracted.

Since the random coincidences in the proton-gated $\gamma$ spectrum
around E$_{\gamma}=7$ MeV is dominated by neutrons, it can be used to
eliminate the neutron-related events from the real coincidences by
subtracting it with a weighting factor, which is defined by the ratio
of the corresponding time windows. In the resulting $p-\gamma$
coincidence spectrum, the peak observed at 8 MeV represents
$\gamma$-rays from the AGDR $\rightarrow$ IAS transition only.

The energy distribution of the $\gamma$ rays was fitted by a Gaussian
curve and a second-order polynomial background as shown in
Fig.~\ref{fig:gamma}.  The obtained energy and width of the transition
are $E_\gamma = 8.090\pm0.013$ MeV and $\Gamma=$2.2 MeV.  However, the
energy calibration of the CACTUS spectrometer has been performed with
photopeaks having significantly smaller width than giant
resonances. In order to determine the real energy of the resonance,
GEANT Monte-Carlo simulations were performed and convoluted with a
Gaussian function with the width of the resonance. This analytical
procedure caused a reduction of 10\% in the position of the peak,
which was taken into account when the final energy of the transition
was extracted. As a result, the transition energy is
$E_\gamma=8.90\pm0.02$ MeV including only the statistical error.

The contribution of the systematical error stems from the uncertainty
of the energy calibration, which is estimated to be $1.0\%$, so the
final transition energy is $E_{AGDR}-E_{IAS}=8.90\pm0.09$ MeV.  The
energy and width of the transition agree well with the previously
measured values of Refs.~\cite{st80,ni85} but having significantly
smaller error bars.
 
\section{Theoretical analysis}
The AGDR and IAS excitation energies are calculated with the 
self-consistent relativistic proton-neutron random-phase
approximation (pn-RRPA) \cite{Paar2003,Paar2004} 
based on the Relativistic Hartree (RH) model \cite{VALR.05}. 
As in our previous studies of the AGDR \cite{kr13,kr13a}, the 
calculation is based on family of
density-dependent meson-exchange (DD-ME) interactions, for which the
constraint on the symmetry energy at saturation density has been
systematically varied: $J =$ 30, 32, 34, 36 and 38 MeV, and the
remaining model parameters have been adjusted to accurately 
reproduce nuclear-matter properties (the saturation density, the 
compression modulus) and the
binding energies and charge radii of a standard set of spherical
nuclei \cite{VNR.03}.  These interactions were also used 
in Ref.~\cite{kl07} to study a possible correlation between the
observed pygmy dipole strength (PDS) in $^{130,132}$Sn and the
corresponding values for the neutron-skin thickness. In addition, the 
relativistic functional DD-ME2 \cite{LNVR.05} will be also used 
in the calculation of the excitation energies of the AGDR with 
respect to the IAS. We note that the relativistic RPA
with the DD-ME2 effective interaction predicts the dipole polarizability
\begin{equation}
\alpha_D = {{8 \pi}\over 9} e^2~m_{-1}
\label{dip-pol}
\end{equation}
(directly proportional to the inverse energy-weighted moment $m_{-1}$)
for $^{208}$Pb: $\alpha_D$=20.8 fm$^3$, in 
agreement with the recently obtained experimental value: $\alpha_D = (20.1\pm 0.6)$
fm$^3$ \cite{ta11}.

\begin{figure}[t]
\centering
\includegraphics[width=80mm]{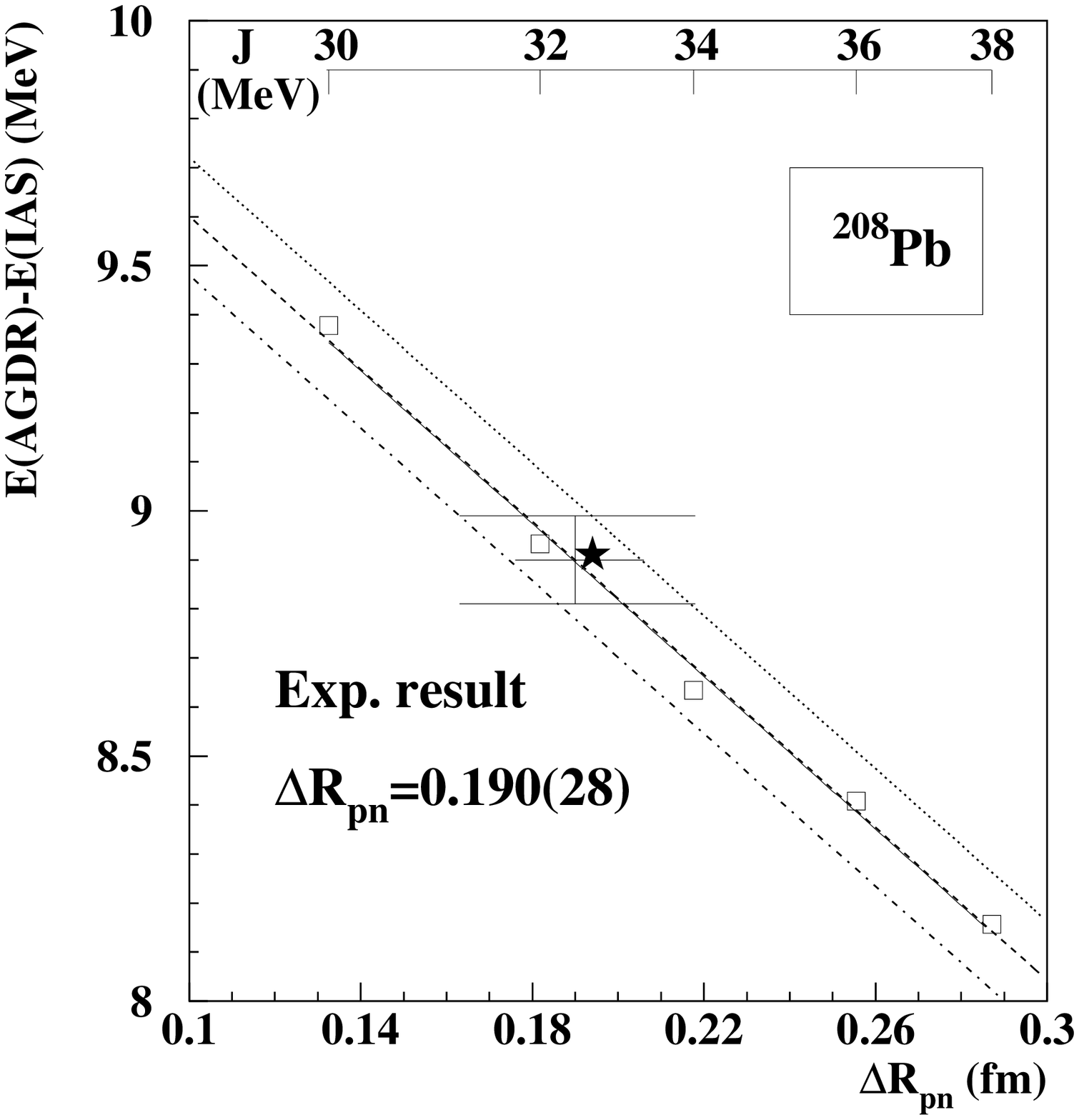}
\caption{The difference in the excitation energies of the AGDR and the
  IAS for the target nucleus $^{208}$Pb, calculated with the pn-RRPA
  using five relativistic effective interactions characterized by the
  symmetry energy at saturation $J =$ 30, 32, 34, 36 and 38 MeV
  (squares), and the interaction DD-ME2 ($J =32.3$ MeV) (star). The
  theoretical values $E(AGDR) - E(IAS)$ are plotted as a function of
  the corresponding g.s. neutron-skin thickness $\Delta R_{pn}$, and
  compared to the experimental value $E(AGDR) - E(IAS) = 8.90 \pm 0.09
  $ MeV.
\label{fig:skin208}}
\end{figure}

The results of the calculations for $^{208}$Pb are shown in
Fig.~\ref{fig:skin208}. The difference in the excitation energies of the
AGDR and the IAS, calculated with the pn-RRPA based on the RH
self-consistent solution for the g.s. of the target nucleus, is
plotted as a function of the corresponding RH predictions for the
neutron-skin thickness. For the excitation energy of the AGDR we take
the centroid of the theoretical strength distribution, calculated in
the energy interval above the IAS that corresponds to the measured
spectrum of $\gamma$-ray energies: E$_{\gamma}$=6 to 14.8 MeV
(Fig.~\ref{fig:gamma}). A single peak is calculated for the IAS. For the
effective interactions with increasing value of the symmetry energy at
saturation $J =$ 30, 32, 34, 36 and 38 MeV (and correspondingly the
slope of the symmetry energy at saturation \cite{VNPM.12}), one notices 
a linear decrease of $E(AGDR) - E(IAS)$ with increasing values 
of the neutron skin $\Delta R_{pn}$.  The value calculated with DD-ME2
($J =32.3 $MeV) is denoted by a star.

The uncertainty of the theoretical predictions for the neutron-skin
thickness is estimated around 10 \%. This uncertainty was 
adopted for the 
differences between the neutron and proton radii
for the nuclei $^{116}$Sn, $^{124}$Sn, and $^{208}$Pb,  when the
parameters of the effective interactions with 
$J =$ 30, 32, 34, 36 and 38 MeV, and DD-ME2 
were adjusted \cite{VNR.03,LNVR.05}.  
These interactions were also used to calculate the electric dipole
polarizability and neutron-skin thickness of $^{208}$Pb, $^{132}$Sn
and $^{48}$Ca, in comparison to the predictions of more than 40
non-relativistic and relativistic mean-field effective interactions
\cite{Pie12}.  From the results presented in that work one can also
assess the accuracy of the present calculation.

From the comparison to the experimental result for $E(AGDR) - E(IAS)$ we
deduce the value of the neutron-skin thickness in $^{208}$Pb: $\Delta
R_{pn} = 0.190 \pm 0.028 $ fm (including the 10\% theoretical
uncertainty).  In Table I this value is compared to previous results
obtained with a variety of experimental methods. 

In parallel with our work the neutron-skin thickness has been
extracted from coherent pion photo-production
cross sections \cite{ta03}. The half-height radius and diffuseness of the
neutron distribution are found to be 6.77$\pm$0.03(stat) fm and
0.55$\pm$0.01(stat)$^{+0.00}_{-0.025}$(sys) fm respectively, corresponding to
a neutron skin thickness $R_{pn}$ =0.19$\pm$0.03(stat)$^{+0.00}_{-0.03}$(sys) fm \cite{ta03}, which agrees very well with our results.

The very good 
agreement with all available data supports the reliability of the method 
employed in the present study.

\begin{table}[htb]
\centering
\caption{\label{tab:table2}The value of the neutron-skin thickness 
of $^{208}$Pb determined
in the present work compared to available data.}

\begin{tabular}{lllc}
\hline\hline
\textrm{Method}& \textrm{Ref.}& \textrm{Date}& \textrm{$\Delta  R_{pn}$} (fm)\\ 
\hline
($p$,$p$) 0.8 GeV & \cite{ho80} & 1980 & 0.14 $\pm$ 0.04\\ 
($p$,$p$) 0.65 GeV & \cite{sta94} & 1994 & 0.20 $\pm$ 0.04 \\ 
($\alpha,\alpha\prime$) IVGDR 120 MeV & \cite{kr94} & 1994 & 0.19 $\pm$ 0.09 \\ 
antiproton absorption & \cite{tr01} & 2001 & 0.18 $\pm$ 0.03 \\ 
($\alpha, \alpha\prime$) IVGDR 200 MeV & \cite{kr04} & 2003 & 0.12 $\pm$ 0.07\\ 
pygmy res.             & \cite{kl07} & 2007 & 0.180 $\pm$ 0.035 \\
pygmy res.             & \cite{ca10} & 2010 & 0.194 $\pm$ 0.024 \\
($\vec p$,$\vec p\prime $) & \cite{ta11} & 2011 & 0.156 $\pm$ 0.025\\ 
parity viol. ($e$,$e$) & \cite{ab12} & 2012 & 0.33 $\pm$ 0.17 \\
AGDR & pres. res. & 2013 &  0.190 $\pm$ 0.028 \\
\hline
\end{tabular}
\end{table}

\section{Constraints on the symmetry energy from the energy difference of the
AGDR and the IAS}

In addition to correlating the excitation energy of the AGDR to
the neutron skin, we have also used the AGDR to determine 
constraints on the symmetry energy
at saturation density (J), and slope of the symmetry energy (L).
Figure 5 shows that the J-L plot is particularly instructive because the
AGDR constraint can be directly compared to those of the dipole
polarizability and the pygmy resonances (PDR). It is 
important to note that constraints from AGDR, $\alpha_D$, and PDR on this
plot are obtained using
the same family of energy density functionals, so one can determine whether 
different excitations probe the same property of the symmetry energy.
From the AGDR analysis, we obtain constraints $J=32.7 \pm 0.6$ MeV and
$L=49.7 \pm 4.4$ MeV.

Fig. 5 also shows a set of J-L constraints from a number of previous studies.
A set of constraints from heavy ion collisions (HIC), within two standard deviations from
the minimum, corresponding to a 95\% confidence level, is 
confined by the two solid lines in the (L, J) plane \cite{ts12}.
The different rectangles in the figure denote the following 
constraints:
from Quantum Monte Carlo (QMC) and neutron star \cite{sg12},
from nuclear binding energies (FRDM) \cite{mm12},
from isobaric analog states (IAS) \cite{dl12,dl09},
from proton elastic scattering ($^{208}$Pb (p,p)) \cite{ze10},
from pygmy dipole resonances (PDR); LAND 2007 \cite{kl07}
and Carbone 2010 \cite{ca10},
from dipole polarizability experiment \cite{ta11} 
and from the present result for the AGDR. The J-L constraints 
from $\alpha_D$ are reanalyzed using the same set of DD-ME 
effective interactions as in the study of AGDR.  One can observe in
figure that the mean values of  J-L parameters obtained from the AGDR
and $\alpha_D$ almost coincide, however, the AGDR provides more
stringent constraints.

\begin{figure}[bht]
\centering
\includegraphics[width=80mm]{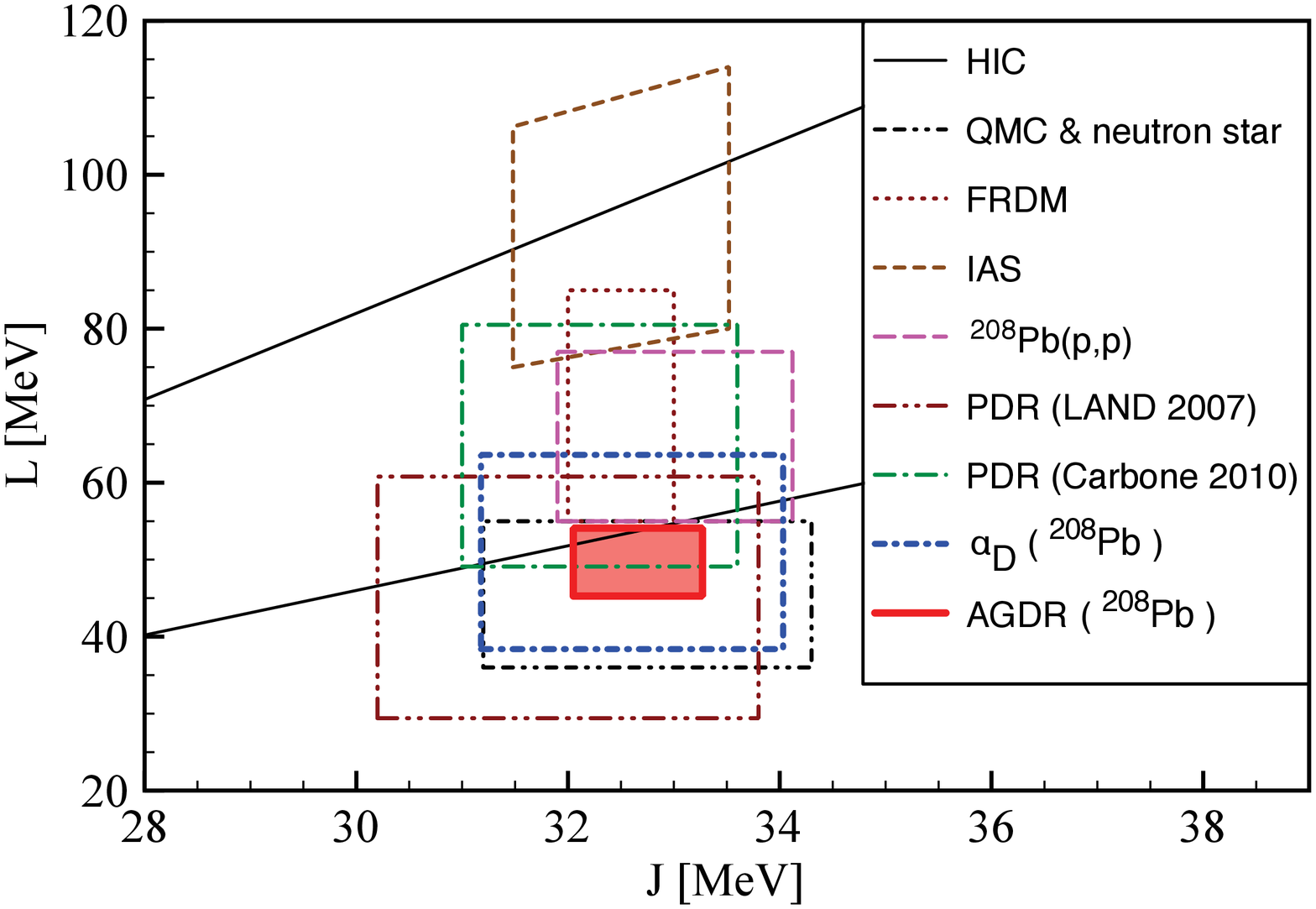}
\caption{Constraints on the slope L and magnitude J of the symmetry energy
at saturation density from different experiments compared to our present
result (AGDR).}
\end{figure}

\section{Conclusions}
%
In this study we have analyzed the $\gamma$ decay of the AGDR to
the IAS excited in the $^{208}$Pb(p,n$\gamma$$\bar p$) $^{207}$Pb
reaction. Using the experimental value obtained for the energy
difference of the AGDR and the IAS, and comparing with the results of 
the RH+pn-RRPA model, we have been able to determine the corresponding 
neutron skin thickness in $^{208}$Pb: $\Delta R_{pn}$= 0.190
$\pm$ 0.028 fm. The agreement between the present result and
values obtained in previous experiments using different methods is very good.  
In particular, the value obtained here is in accordance with results of a very recent high-resolution
study of electric dipole polarizability $\alpha_D$ in $^{208}$Pb
\cite{ta11},  the correlation analysis of $\alpha_D$ and $\Delta
R_{pn}$ \cite{Pie12}, as well as with the Pb Radius Experiment (PREX) that 
used parity-violating elastic electron scattering at JLAB \cite{ab12}.

The measured energy difference between the AGDR and the IAS has also been 
used to constrain possible values of the symmetry energy
at saturation density (J), and the slope of the symmetry energy (L). 
We have found good agreement between constraints that result from 
the AGDR and $\alpha_D$, whereas the discrepancy with the constraint 
obtained from the pygmy resonance is probably due to the missing strength in PDR 
experiments \cite{pa10}. Therefore, measurements of the
AGDR might be important not only to constrain possible values of J and L, 
but also to understand differences between results obtained in various
experiments. Since the mean values of J-L constraints obtained from the AGDR
and $\alpha_D$ appear in excellent agreement, obviously the two very different
collective modes of excitation in nuclei probe the same underlying physical
content. The main advantage of the method based on the AGDR
compared to the $\alpha_D$ analysis and most of the previous studies is 
that it provides more stringent constraints on the symmetry energy parameters.


\section{Acknowledgments}

This work has been supported by the Hungarian OTKA Foundation No.\,
K106035. This research was also supported by the
European Union and the State of Hungary, co-financed by the European 
Social Fund in the framework of T\'AMOP-4.2.2/B-10/1-2010-0024 and
T\'AMOP 4.2.4.A/2-11-1-2012-0001  `National Excellence Program'.

\end{document}